\documentclass[a4paper]{jpconf}
\usepackage{graphicx}
\usepackage{rotating}
\usepackage{amsmath}
\usepackage{amssymb}
\usepackage{float}
\usepackage{hyperref}
\begin{document}

\def\chan{{\sl CXO}}
\def\xmm{{\sl XMM}-Newton}
\newcommand{\hst}{{\sl HST}}
\newcommand\apj{ApJ}
\newcommand\pasj{PASJ}
\newcommand\apjs{ApJS}
\newcommand\mnras{MNRAS}
\newcommand\aap{A\&A}
\newcommand\nat{Nature}
\newcommand\apjl{ApJL}
\newcommand\ssr{SSR}
\newcommand\aplett{Astrophysical Letters}
\newcommand\araa{ARAA}
\newcommand\physrep{Physics Reports}

\title{Toward understanding the physical underpinnings of spatial and spectral morphologies of pulsar wind nebulae}

\author{O Kargaltsev$^1$,  N Klingler$^1$, S Chastain$^1$ and G G Pavlov$^2$}

\address{$^1$ Department of Physics, The George Washington University, Washington, DC 20052, USA}
\address{$^2$ Department of Astronomy \& Astrophysics, Pennsylvania State University, University Park, PA 16802, USA}

\ead{kargaltsev@gwu.edu}

\begin{abstract}
We discuss the observational properties of pulsar wind nebulae (PWNe) linking them to the injected (at the termination shock) electron spectral energy distribution and parameters of pulsar magnetospheres. 
In particular, we (1) present spatially-resolved  {\sl Chandra} ACIS spectral maps of twelve PWNe and measure the slopes of the uncooled PWN spectra just downstream of the termination shock obtained from these maps,  
 (2) consider the connections between PWN morphologies and predictions of the  magnetospheric  emission models  
  and (3) discuss the limits on the maximum energies of particles
   in PWNe from X-ray and $\gamma$-ray observations. 
\end{abstract}

\section{Introduction}
Pulsars are among nature's most powerful 
 particle accelerators, capable of producing and accelerating particles up to PeV energies.  
For rotation-powered pulsars (RPPs), most of the neutron star's (NS's) rotational energy is imparted into a magnetized relativistic pulsar wind (PW) which emits nonthermal radiation from radio to TeV $\gamma$-rays, forming a pulsar wind nebula (PWN).
  The particle wind is ultra-relativistic immediately beyond the  pulsar magnetosphere (light cylinder), but the interaction with the surrounding medium causes the bulk flow to slow down abruptly  at a termination shock (TS). 
 The shocked magnetized  PW is confined between the termination shock (TS) and contact discontinuity (CD). 
  It produces synchrotron radiation observed from radio to GeV $\gamma$-rays and inverse Compton radiation observed from GeV to TeV energies (see, e.g., reviews  \cite{Gaensler2006,Kargaltsev2015,Reynolds2017,2017arXiv170309311S}).

The morphologies and spectra of PWNe depend on the anisotropy of the wind, its magnetization, particle acceleration efficiency, magnetic field strength and pulsar velocity. 
 While most pulsars have reliably measured spin-down energy loss rates ($\dot{E}$), it is much more difficult to determine the other parameters responsible for the diverse morphologies, radiative efficiencies and spectra of PWNe.  It is natural to assume that the angle $\zeta$ between the pulsar's spin axis and the observer's line of sight 
  and the inclination angle $\alpha$ between the spin and magnetic dipole axes (if the field is indeed largely dipolar) would leave an imprint on the PWN properties. 
 However, these angles are notoriously difficult to measure. 
Constraints on these angles can be obtained by comparing theoretical models of pulsar magnetospheric emission with the {\sl Fermi} LAT GeV light curves and radio light curves (e.g., \cite{Watters2009}).
 However, different magnetosphere models can lead to differing predictions. 
The PWN spectra are also affected by cooling, which is expected to become progressively more important with increasing distance from the pulsar. 
Therefore, to determine the particle spectral energy distribution (SED) injected at the TS, one must obtain spatially-resolved PWN spectra.

For a stationary or slow-moving pulsar, the CD (and therefore the PWN) will expand as more energy and particles are injected by the pulsar, until the radiative and adiabatic expansion losses balance the rate of energy input.
For an isotropic PW, one would expect the CD/PWN to have a spherical shape; however, no spherical PWNe are observed which shows that  pulsar winds are intrinsically anisotropic. 
Many PWNe exhibit prominent equatorial and polar components (e.g., tori and jets) which reflect symmetry with respect to the pulsar spin axis (see, e.g., the Crab and Vela PWNe in figure~1).  
In such cases it becomes possible to measure  the angle between the line of sight and the pulsar's spin axis and compare it with the predictions of magnetosphere emission models.
Some PWNe display  nearly-axial symmetry, but identifying the equatorial and polar outflows can be challenging (e.g., G21.5--0.9 and MSH 11--62 in figure~1).
The anisotropic winds can
  lead to anisotropic PWN spectra, flow speeds, magnetic fields, and cooling trends, requiring high-resolution imaging, spatially-resolved spectroscopy, and spatially-dependent models accounting for the wind anisotropy. 

It is even more challenging to comprehend the effects of the magnetic inclination 
   $\alpha$ on PWN parameters.  
One can hypothesize that lower values of $\alpha$ would result in a  highly magnetized wind with a stronger polar outflow (compared to the equatorial torus) and also perhaps a reduced particle acceleration efficiency and lower PWN luminosities (see modeling by \cite{Buhler2016}).

The  ``natal kicks'' that pulsars receive in 
  supernova explosions 
 result in large pulsar velocities with the average 3D velocity  $v_{\rm psr}\sim400$ km s$^{-1}$ \cite{Hobbs2005}.
 This implies that most pulsars  remain inside their host SNRs only for a few tens of kyrs; after 
  that they travel through the interstellar medium (ISM) where the speed of sound $c_{\rm ISM}\ll v_{\rm psr}$ ($c_{\rm ISM}\sim3-30$ km s$^{-1}$, depending on the ISM phase).
In this environment, the pulsar motion  becomes supersonic (Mach number $\mathcal{M}\equiv v_{\rm psr}/c_{\rm ISM}> 1$), and the ambient medium ram pressure strongly modifies the PWN appearance  
leading to  formation of extended pulsar tails (see \cite{Reynolds2017,Kargaltsev2017} 
 for reviews on supersonic PWNe).
Although in supersonic PWNe (SPWNe) the structures formed by the anisotropic post-TS wind (e.g., jets and tori) are 
 deformed by the ram pressure, 
 in some cases the torus-jet structure can still be identified in high-resolution images (see, e.g., J1509--5850, Geminga, B1706--44, and B0355+54 in figure~2 of \cite{Kargaltsev2017}).

\begin{figure}
\centerline{\includegraphics[scale=0.07]{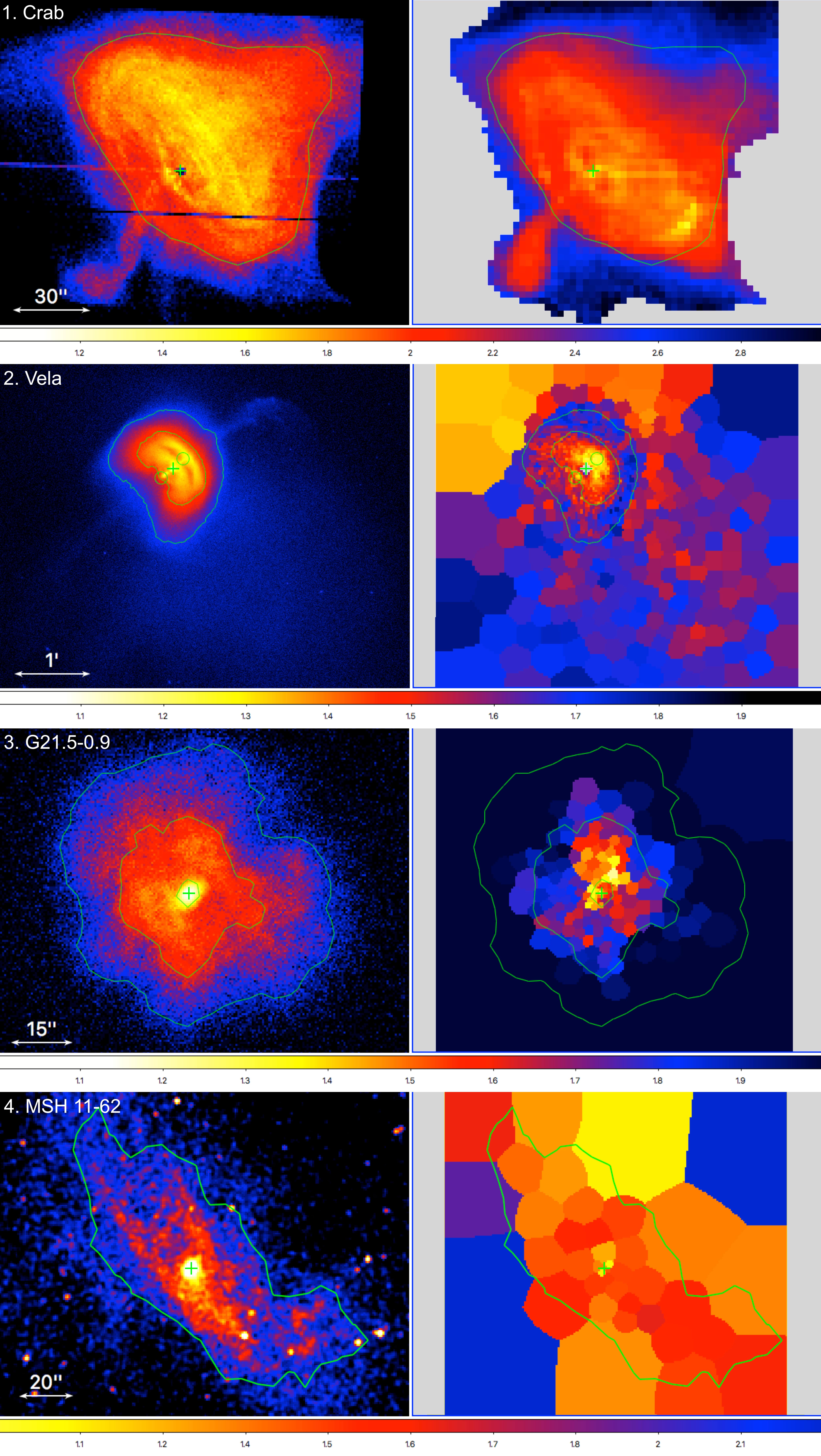}}
\vspace{-0.4cm}
\caption{{\sl Left panels: } {\em CXO} ACIS images (in 0.5--8 keV) of bright, well-resolved PWNe  from table 1.  {\sl Right panels:} Adaptively-binned spatially-resolved spectral maps for the PWNe shown on the left.  The color bars represent the photon index $\Gamma$ measured in the 0.5--8 keV band.  The green contours are shown for illustrative purposes, and the green crosses mark the pulsar positions.  The gray-colored areas in panels 1 and 2 were excluded from mapping.  The Crab spectral map is adopted from \cite{Mori2004}.}
 \label{fig:spmaps1}
\end{figure}

\begin{figure}
\centerline{\includegraphics[scale=0.14]{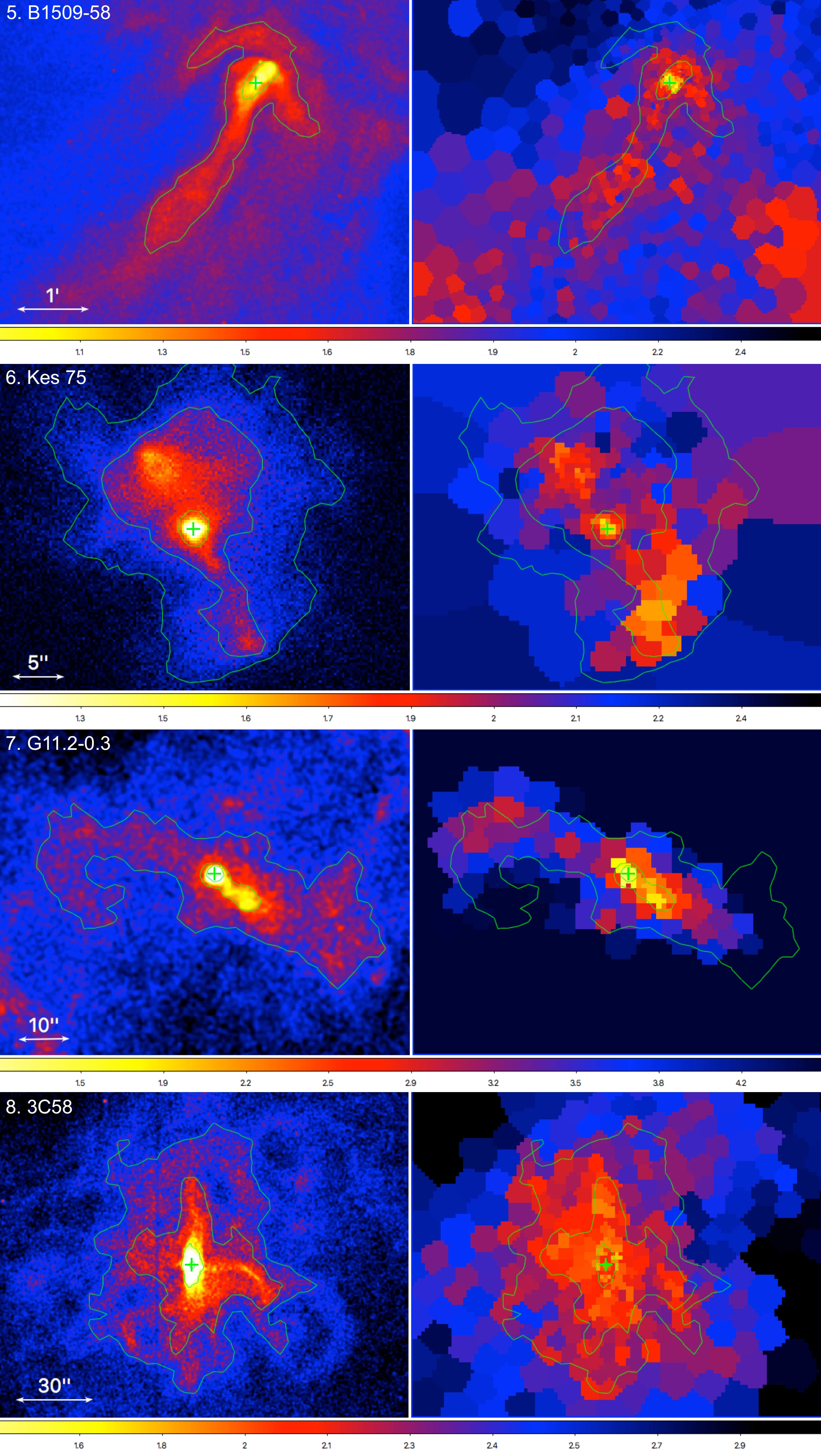}}
\vspace{-0.4cm}
\caption{ ACIS images and spectral maps of bright, well-resolved PWNe  (continued from figure 1).}
\label{fig:spmaps2}
\end{figure}

\begin{figure}
\centerline{\includegraphics[scale=0.14]{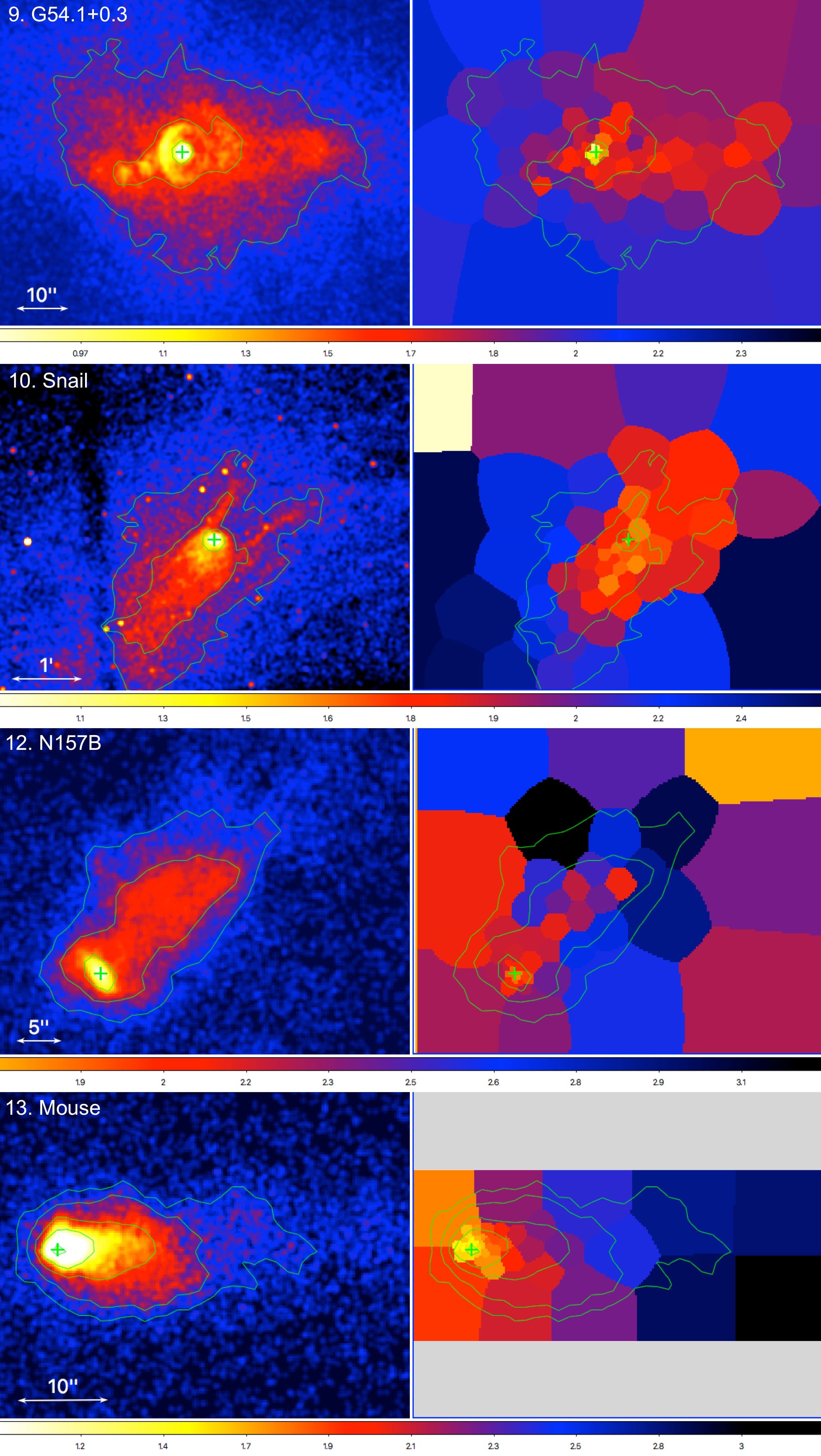}}
\vspace{-0.4cm}
\caption{ ACIS images and spectral maps of bright, well-resolved PWNe (continued from figures 1 and 2).}
\label{fig:spmaps3}
\end{figure}

  Great progress in measuring the physical properties of PWNe has been made with the Advanced CCD Imaging Spectrometer (ACIS) onboard the  \textit{Chandra X-ray Observatory} (\textit{CXO}), whose unprecedented angular resolution combined with the low ACIS background 
   makes it possible to resolve (spatially and spectrally) the complex anisotropic structures of PWNe.  
On smaller scales ($\sim$0.01--0.1 pc), structures such as bow shocks, wisps, knots, rings, jets, arcs and combinations thereof have been observed (see figures~\ref{fig:spmaps1}--\ref{fig:spmaps3}).  
On larger scales ($\sim$0.1--10 pc), structures such as tori and long diffuse tails directed opposite the pulsar's motion are often seen, and in a few cases (associated with particularly fast-moving pulsars), puzzling linear structures strongly misaligned with the pulsar's direction of motion have been discovered (see, e.g., figures~2 and 9 in \cite{Kargaltsev2017}).
 Many PWNe include various combinations of the above structures but sometimes can have amorphous morphologies.

In this paper we focus on  investigating (1) the slopes of the particle SEDs injected at the TS using adaptively binned spectral maps, (2) the correspondences between the angles $\alpha$ and $\zeta$ inferred from PWN morphologies  with those predicted by  magnetospheric emission models and the possible dependencies of PWN parameters on these angles and (3) the constraints on maximum  PW particle energies.

\section{Spatially-resolved X-ray Spectra of PWNe}
 
Over the last 18 years of its operation, the {\sl CXO} has accumulated a substantial amount of ACIS data on bright, well-resolved PWNe with different morphologies. 
At  present, there are over 20 sufficiently bright PWNe with  deep  ACIS exposures  (see table 1), suitable for detailed spatially-resolved spectroscopy.  For the brightest of these PWNe (shown in figures 1--3), this can be done by producing adaptively-binned spectral maps with spatial resolutions ranging from a few arc seconds to about an arc minute.

\begin{table*}
\scriptsize
\caption{\scriptsize {\sl Chandra} ACIS observations of bright X-ray PWNe suitable for spatially resolved spectroscopy.  Column 3 lists the on-axis exposure times of imaging ACIS observations 
  (the number of separate observations is in parentheses).  Column 4 gives the total observed (absorbed) X-ray flux (0.5--8 keV).  Column 5 lists prominent features found in the PWN.  The first group of PWNe (above the horizontal line) is
    located inside SNRs, the second group is supersonic PWNe. $^{\dagger}$ -- the scientific exposure for the Crab PWN is much smaller (107 ks).}
\centering
\setlength{\tabcolsep}{0.3pt}
\begin{tabular}{lcccc}
\hline\noalign{\smallskip}
Number & Name & Exposure (ks)\ \ \ \ \ \ &  $F_{\rm X}$ ($10^{-11}$ cgs) & Features  \\ [3pt]
\hline
1 & Crab & 565$^{\dagger}$ (57) & 2000 & ring, torus, jets \\                                
2 & Vela & 630 (22) & 6.1 &  arcs, jets \\                                     
3 & G21.5--0.9 (J1833-1034) & 738 (82) & 3.8  & tori, jets \\                                 
4 & MSH 11--62 & 472 (9)  & 3  &  torus?, 2 arcs(?)\\                          
5 & G320.4--1.2 (B1509--58) & 331 (10) & 5.3 &  arcs, jet, tail(?) \\                
6 & Kes 75 (J1846-0258) & 383 (8) & 1.5 & jets, torus \\                         
7 & G11.2--0.3 (J1811-1925) & 480 (12) & 0.6 & jets, torus(?) \\                             
8 & 3C58 (J0205+6449) & 392 (4) & 0.7 & jets, loops, torus \\                                 
9 & G54.1+0.3 (PSR J1930+1852) & 326 (5) & 0.5 & torus, jets \\                                
10 &  Snail/G327.1--1.1 & 386 (4)  & 0.3 & tail, prongs     \\                    
11 & MSH 15--56 (G326.3--1.8) & 56 (1) & 0.83 & amorphous morphology \\        
12 & N157B (J0537--6910) & 48 (1) & 0.6 & torus, tail/trail \\                     
\hline
13 &  Mouse (J1747--2958) & 154 (5) & 0.7 &  equatorial outflow, tail \\                     
14 &  Lighthouse (J1101--6101) & 302 (7) & 0.2 & misaligned outflow \\        
15 & Geminga & 680 (14) & 0.075 &  equatorial \& lateral outflows \\                
16 & CTB 80 (B1951+32) & 85 (1) & 0.6 & torus, tail(?) \\                                 
17 & J1509--5850 & 413 (5) & 0.05 & bow shock, jets, tail \\            
18 & Mushroom (B0355+54) & 461 (9) & 0.03 & equatorial outflow, jets, tail \\                    
19 & J1741--2054 & 331 (7) & 0.03 & bow shock, tail \\                         
20 & J0357+3205 & 134 (4) & 0.04 & (disconnected) tail \\                    
21 & Guitar (B2224+65) & 195 (6) & 0.006 & bow shock, misaligned outflow \\                       
\noalign{\smallskip}\hline
\end{tabular}\\
\vspace{-0.3cm}
\label{tbl-pulsars}
\end{table*}

A number of previous studies have investigated the spectral properties of PWNe observed by the \chan\  (\cite{Kargaltsev2013}
 and references therein). 
 However, even for relatively bright, extended PWNe, most authors report spatially-averaged spectral indices from differently defined regions (e.g.,  based on morphology, brightness contours, number of counts, etc).
 This often does not allow one to cleanly separate the effects of varying injection spectra (measurable just downstream the TS)
 from cooling trends or to look for azimuthal spectral dependences associated with the polar and equatorial outflows. 
The lack of uniformity also complicates the comparison between different PWNe.   
An example of complications arising from the measurements of spatially-averaged spectra can be seen in the analysis of deep observations of the Mouse PWN (Klingler et al.,  in prep.). The high-S/N spectrum of the fairly compact ($50''$-long) X-ray tail can be fit well by a power-law  (PL) model  with photon index $\Gamma=2.09\pm0.03$. 
Yet, our spatially-resolved analysis shows that the spectral slope varies dramatically with distance from the pulsar along the same region, with $\Gamma$ increasing from $1.65\pm0.06$ to $3.0\pm0.1$. 
Thus,  to study the dynamics of pulsar winds and the particle acceleration mechanisms, one needs to create accurate, high-resolution spectral maps following a uniform approach.

  An adaptive binning methods using weighted Voronoi tessellations (WVTs)
   were developed in \cite{Cappellari2003,Diehl2006} to analyze galaxy cluster emission. 
We adopted the technique to PWNe and incorporated a method to utilize an arbitrary number of 
 ACIS observations by applying observation-specific calibration corrections, calculating region- and observation-specific detector responses, and performing joint fitting of spectra extracted from the individual observations for each spatial bin of the spectral map.
   
These spectral maps can be used to  (1)  measure the slope of the injected electron SED, 
(2) characterize and visualize synchrotron cooling trends\footnote{The characteristic cooling length scales depend on the bulk flow velocity, the magnetic field strength, the presence of in-situ acceleration, the geometry of the outflow and ISM entrainment.} which vary among PWNe and, possibly, among  different features of the same PWN (e.g., jets vs.\  tori),  (3) constrain the flow velocities, magnetic fields and particle diffusion,   
(4) search for in-situ acceleration sites located downstream of the termination shock (e.g., due to magnetic turbulence and reconnection),  
(5) understand the origin of the PWN features (e.g., puzzling asymmetric structures seen in several PWNe associated with fast-moving pulsars).  
 Recently, 
 Porth et al.\ 
  \cite{Porth2016} presented calculations of spatially-resolved, radial (i.e.,\ azimuthally-averaged) spectra which take into account particle cooling, advection and diffusion using 3D MHD. 
The results suggest that the competition between the advection and diffusion 
   can result in different spectral properties of PWNe with different magnetic fields and ages. The latitudinal dependence  of the diffusion transport can cause anisotropy in the particle SED  within the same PWN (e.g., toroidal vs.\ polar directions) which has not yet been investigated.  
 Once synthetic spatially-resolved spectra are calculated for 3D models, they  can be compared to the  spectral maps to obtain further constraints on the diffusion coefficients, flow speeds and magnetic field strengths of PWNe.

\begin{figure}
\centerline{\includegraphics[scale=0.50]{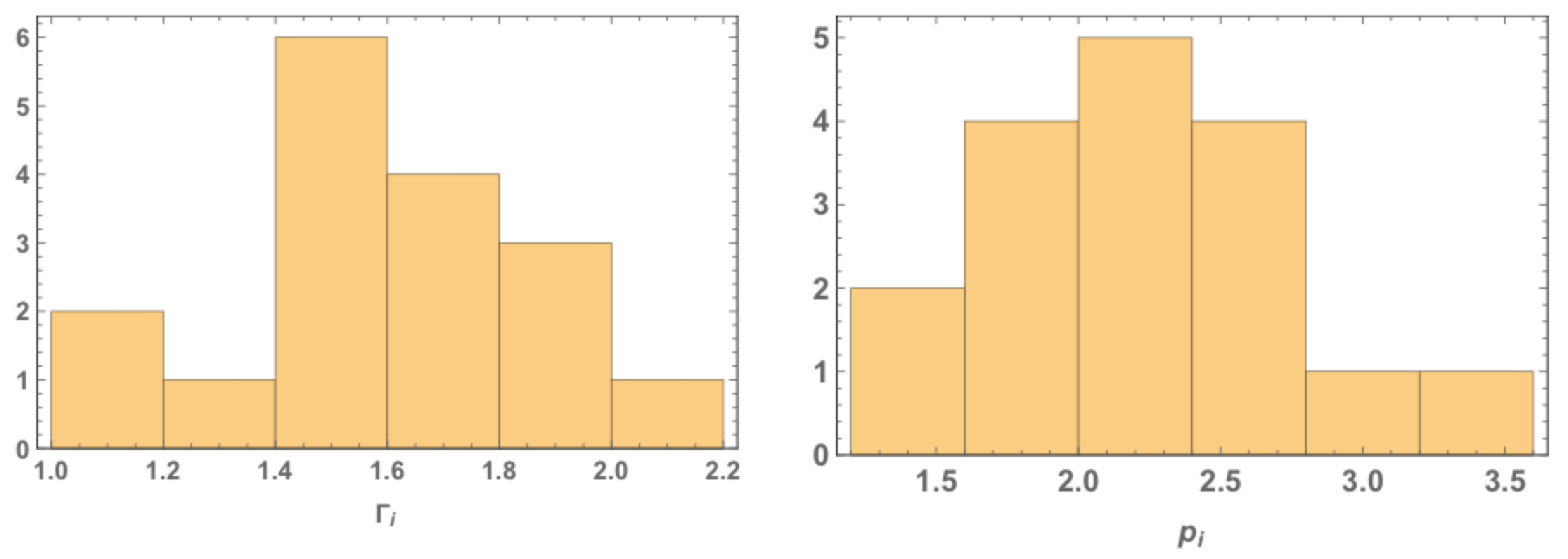}}
\vspace{-0.4cm}
\caption{Histograms of $\Gamma_{\rm i}$ and $p_{\rm i}=2\Gamma_{\rm i}-1$ for the 17 PWNe listed in table 2 (note that pulsars 11 and 14 are omitted from table 2; see the caption).
 The bin width corresponds to the average measurement uncertainty.}
\label{fig:histo}
\end{figure}

The spectral maps shown in figures 1--3 (right panels) reveal varying injection spectra  (see table \ref{tbl-inj-spectra}), drastically different cooling trends, and other interesting features.
 For example, the Vela PWN map offers two surprises. 
On a large scale, there appears to be an enveloping region of spectral hardening outside  the inner nebula (in all directions but the south; see figure~\ref{fig:spmaps1}).
On small scales (within the inner compact nebula) the southeast inner jet appears to be brighter than the northwest jet suggesting that the former jet is approaching while the latter one is receding \cite{Helfand2001,Durant2013}.
 Interestingly, the spectra of the jets appear to differ, with the fainter northwest inner jet appearing harder ($\Gamma=1.25\pm0.02$) than the brighter southeast inner jet  ($\Gamma=1.36\pm0.04$).  
Although Doppler boosting can change the apparent brightnesses of jets, it should not affect the jets' spectra.
Moreover, in the G11.2--0.3 and G54.1+0.3 maps, the apparently approaching brighter jets 
   appear to be harder.  The map of 3C58 shows that the edge-on torus and western jet are the hardest features of the PWN embedded into a softer emission. Therefore, any averaged radial  profiles of photon index will not reflect the actual anisotropic dependences.     
 Overall, in the cases when both polar (jets) and equatorial (torus) components can be identified, their spectra appear to be similar. In the context of the Komissarov \& Lyubarsky 
 model \cite{Komissarov2004} where the jets form from the equatorial outflow diverted by magnetic field hoop stress, it means that the particle SED does not change in this process.
  In G21.5--0.9, the PWN brightness distribution appears roughly symmetric along a line running northeast to southwest, but the spectral map reveals a bulbous lobe with a harder spectrum extending northwest from the pulsar, which reveals the otherwise invisible 
 additional anisotropy  of the PW (figure~\ref{fig:spmaps2}).
The distribution of the photon indices $\Gamma_i$ measured from the spectral maps for the innermost regions of 17 PWNe and the SED slopes  inferred from them ($p_{\rm i}=2\Gamma_i-1$, assuming negligible cooling\footnote{This may not hold for PWNe at large distances (e.g., in the LMC) where small angular scales correspond to large distances.}) is shown in figure \ref{fig:histo}.  
The $p_i$ values span the range from 1.4 (for Vela and G320.4--1.2) to 3.3 (in N157B). 
However, the latter PWN is very remote (it is in the LMC), and the outflow can be affected by cooling even for the smallest region resolvable by  {\sl CXO}. 
  Therefore, a more plausible range of $p_i=1.8$--$2.5$   with a maximum of the distribution between 2.0 and 2.4 (see figure~4). 
We have also looked for  correlations between $\Gamma_i$ and luminosity ($L_X$) or radiative efficiency ($\eta_X=L_X/\dot{E}$) of the innermost regions of the brightest PWNe (taken from table 2 of \cite{Kargaltsev2008}). 
Despite the significant scatter, there appears to be a hint of a positive correlation between $\Gamma_i$ and  $L_X$, suggesting that PWNe with harder injected  SEDs  are less luminous (see figure~5, left panel).  In the $\Gamma_i$--$\eta_X$ plot (figure~5, right panel)  the correlation is not seen, but there seems to be a deficit of objects with high efficiencies and hard spectra. Some of the PWNe (e.g., the Mouse, Klingler et al., in prep.; the Lighthouse \cite{Pavan2016};
 N157B \cite{Chen2006})
   show strong cooling trends (the spectra is softening rapidly with increasing distance from the pulsar) while other PWNe (e.g., those associated with PSRs J1509--5850 and B0355+54) do not show any substantial cooling along their long tails (see \cite{Klingler2016a,Klingler2016b}
 for spatially-resolved spectral measurements). Interestingly, the misaligned outflows (see \cite{Kargaltsev2017}
 and also below) do not show   spectral softening with distance from the pulsar (see  figure~7 of \cite{Pavan2016}
  for  the adaptively binned spectral map of the Lighthouse PWN).

\begin{table*}
\scriptsize
\caption{\scriptsize Injection spectra of selected bright PWNe, obtained from spatially-resolved spectral mapping (objects above the horizontal line; shown in figures~1--3) and spatially-resolved spectroscopy (below the horizontal line).  Note that four PWNe (MSH 11--62, the Lighthouse nebula, J0357+3205, and B2224+65) listed in table \ref{tbl-pulsars} were excluded, as the injection spectrum can not be reliably determined in those cases.  Here, we define $p_i\equiv 2\Gamma_i-1$, assuming an uncooled electron SED described by a PL with a slope $p_i$.
 }
\centering
\setlength{\tabcolsep}{0.3pt}
\begin{tabular}{ccccc}
\hline\noalign{\smallskip}
\ \ \ Number \ \ \ & Name & \ \ \ \ \ \ \ \ $\Gamma_{ i}$ \ \ \ \ \ \ \ \ \ & \ \ \ \ \ \ \ \ \ \ $p_i$ \ \ \ \ \ \ \ \ \  \\ [3pt]
\hline
1 & Crab & $1.80\pm0.05$ & $2.60\pm0.10$ \\
2 & Vela & $1.33\pm0.06$ & $1.67\pm0.12$ \\
3 & G21.5--0.9 & $1.43\pm0.09$ & $1.86\pm0.18$ \\
4 & MSH 11--62 & $1.10\pm0.05$ & $1.20\pm0.10$ \\
5 & B1509--58 (G320.4--1.2) & $1.19\pm0.13$ & $1.38\pm0.27$ \\
6 & Kes 75 (J1846--0258) &  $1.85\pm0.10$ & $2.70\pm0.20$ \\
7 & G11.2--0.3 & $1.75\pm0.1$ & $2.79\pm0.46$ \\
8 & 3C58 & $1.97\pm0.07$ & $2.94\pm0.15$ \\
9 & G54.1+0.3 & $1.7\pm0.1$ & $2.30\pm0.21$ \\
10 &  Snail (G327.1--1.1) & $1.66\pm0.06$ & $2.32\pm0.11$ \\
12 & N157B (J0537--6910) & $2.17\pm0.11$ & $3.34\pm0.23$ \\
\hline
13 &  Mouse (J1747--2958) & $1.57\pm0.06$ & $2.14\pm0.12$ \\
15 & Geminga & $1.4\pm0.1$ & $1.8\pm0.2$ \\
16 & CTB 80 & $1.7\pm0.1$ & $2.4\pm0.2$ \\
17 & J1509--5850 & $1.43\pm0.18$ & $1.86\pm0.36$ \\
18 & Mushroom (B0355+54) & $1.54\pm0.05$ & $2.08\pm0.10$ \\
19 & J1741--2054 & $1.5\pm0.15$ & $2.0\pm0.3$ \\
\noalign{\smallskip}\hline
\end{tabular}\\
\vspace{-0.3cm}
\label{tbl-inj-spectra}
 \end{table*}

\begin{table*}
\scriptsize
\caption{\scriptsize Viewing angles $\zeta$ and magnetic inclination angles $\alpha$ for $\gamma$-ray pulsars estimated in different ways:  
$\zeta_{\rm PWN}$ -- obtained from fitting of the PWN geometry in \cite{Ng2004,Ng2008,NgKes75,YatsuB1509} (unless marked with $^\dagger$, in which case it was estimated by us based on the appearance of the tori/jets);   
 $\alpha_{\rm pred}$ -- the predicted value of $\alpha$ based on $\zeta_{\rm PWN}$ from  figure~2 of \cite{Watters2009}
 for the two pole caustic (TPC) and outer gap (OG) magnetosphere emission models;
$\zeta_{\rm mod}$ and $\alpha_{\rm mod}$ --  allowed values of $\zeta$ and $\alpha$ obtained from figure~2 of \cite{Watters2009}
  for the pulsar $\gamma$-ray and radio light curve properties taken from the 2nd Fermi Pulsar Catalog \cite{Abdo2013}.
  The object numbers correspond to those in table 1, and the objects first presented in this table (PSR B1706--44 and the Dragonfly PWN with PSR J2021+3651) are numbered sequentially (starting with 22).  
 $^a$ -- The OG model simulations can not produce two $\gamma$-ray peaks with a phase separation $>0.5$ and the lack of a radio pulse.
$^b$ -- The model can not produce the observed light curves for the inferred $\zeta_{\rm PWN}$, so we list the $\alpha$ values for the $\zeta$ value closest to the observed one (which we note in parentheses in the subscript of $\zeta$). 
$^\ddagger$ -- Indicates pulsars not seen in radio.
 }
\centering
\setlength{\tabcolsep}{0.2pt}
\begin{tabular}{llccccccc}
\hline\noalign{\smallskip}
\# \ \ \  &  Name  & \ \ \ \ \ \ $\zeta_{\rm PWN}$ \ \ \ \ \ \  & \ \ \ $\alpha_{\rm pred,TPC}$ \ \ \ 
& \ \ \ $\alpha_{\rm pred,OG}$ \ \ \ & \ \ \ \ $\zeta_{\rm mod,TPC}$ \ \ \ \  & \ \ \ \ $\zeta_{\rm mod,OG}$ \ \ \ \ & \ \ \ $\alpha_{\rm mod,TPC}$ \ \ \ &  \ \ \ $\alpha_{\rm mod,OG}$ \ \ \ \\ [3pt]
\hline
1  &  Crab   &  $61.3\pm1.2$  &  63-86  &   78-90 ($\zeta_{68}$)$^b$  &  45-90  &  68-90  &  40-90  &  55-90  \\
2  &  Vela   &  $63.6\pm0.7$  &  61-86  &   78-90 ($\zeta_{68}$)$^b$  &  45-90  &  68-90  &  40-90  &  55-90  \\
3  &  G21.5-0.9  &  $85.4_{-0.3}^{+0.2}$  &  60-90  &  60-90  &  45-90  &  68-90  &  40-90  &  55-90  \\
5  &  B1509-58   &  $\approx$50  &  27-54  &  51-70  &  22-60  &  38-69  &  15-60  &  42-90  \\
6  &  Kes 75$^\ddagger$  &  $62\pm5$  &  10-25  &  18-25 ($\zeta_{70}$)$^b$  &  15-27; 38-90  &  19-31; 68-90  &  2-26; 42-74  &  0-32; 60-90  \\
8  &  3C58   &  $91.6\pm2.7$  &  72 ($\zeta_{75}$)$^b$  &  --$^a$  &  47-75  &  --$^a$  &  42-48; 72-83  &  --$^a$  \\
13  &  Mouse  &  $\sim$70$^\dagger$  &  41-57  &  50-67  &  40-71  &  58-78  &  39-70  &  47-80  \\
15  &  Geminga$^\ddagger$  &  $\sim$90$^\dagger$  &  36-46  &  13-55  &  13-26; 42-62; 85-90  &  76-90  &  35-45; 80-90  &  6-55  \\
17  &  J1509--5850  &  $\sim$90$^\dagger$  &   41 ($\zeta_{62}$)$^b$  &   43-53 ($\zeta_{70}$)$^b$  &  36-62  &  50-71  &  40-65  &  41-61  \\
22  &  B1706--44 &  $53.3\pm4.5$  &  47-54  &  54-60  &  36-62  &  50-71  &  40-65  &  41-61  \\
23  &  Dragonfly   &  $79\pm3$  &  53-90  &  60-90  &  45-90  &  68-90  &  40-90  &  55-90  \\
\noalign{\smallskip}\hline
\end{tabular}\\
 \label{tbl-angles}
\vspace{-0.3cm}
\end{table*}

\section{Connection Between PWN Morphologies and Pulsar Light Curves}
It is natural to expect that the 
  magnetic inclination angle $\alpha$ (between the spin and magnetic dipole axes)  and the viewing angle $\zeta$ 
  (between the line of sight and the spin axis) 
  will leave an imprint on both the pulsar light curves and the compact PWN morphologies. 
The viewing angle can be inferred from the PWN morphology if the PWN has an identifiable torus or ring(s) associated with the TS or, to a lesser extent, if there are two jets of unequal brightness. 
In the former case, the ellipticity of the torus or ring(s) provides a direct measurement of $\zeta$ while in the latter case, the ratio of the jet brightnesses can be related to the $\zeta$-dependent Doppler factor as $f_b = [(1+\beta\cos\zeta)/(1-\beta\cos\zeta)]^{\Gamma+2}$ where $\beta=v_{\rm flow}/c$  \cite{Ng2004}.
 However, the Doppler factor depends on the flow speed, $v_{\rm flow}$, which has been directly\footnote{If the torus inclination is known and it is Doppler brightened on one side, the contrast with the other side can be used to estimate the flow speed in the equatorial region.} measured only in a few cases (Vela PWN outer jet: 0.3$c$--0.6$c$, \cite{Pavlov2003};
 Crab PWN torus: 0.25$c$--0.55$c$, \cite{Bietenholz2001,Hester2008};
 J1509--5850 PWN jets: possibly 0.2$c$--0.3$c$, \cite{Klingler2016b}).
 Therefore, unlike the torus ellipticity fitting, any estimates based purely on the Doppler brightening  are highly uncertain. 
 Inferring $\alpha$ from PWN morphologies is even harder, but some constraints can still be obtained 
 if the prediction  that PWNe with pronounced toroidal components and weak polar outflows (jets) are likely to be associated with pulsars having large $\alpha$ \cite{Buhler2016}
 finds observational support. 
    
Models of pulsar magnetospheric emission (e.g., \cite{Romani1996,Muslimov2003,Watters2009,Romani2010,Kalapotharakos2014,Pierbattista2015,Harding2016})
 provide solutions for $\alpha$ and $\zeta$; however, the predicted values of these angles are often not well constrained, and widely separated solutions are possible.
Moreover, 
  competing models 
    provide different predictions (below we only compare the Two Pole Caustic (TPC) and Outer Gap (OG) magnetosphere emission models -- see \cite{Harding2016}
 for a review).  
 Table 3 lists the values suggested by the PWN morphologies ($\zeta_{\rm PWN}$), and the values ($\alpha_{\rm mod}$ and $\zeta_{\rm mod}$)  predicted by the pulsar light curve modeling (from \cite{Watters2009,Romani2010}).
  To determine plausible ranges of $\zeta_{\rm PWN}$ from PWN morphologies, we used the values from \cite{Ng2004,Ng2008}
 obtained by careful fitting of tori geometry (whenever available), and estimated plausible ranges of $\zeta_{\rm PWN}$ from the appearance of the tori/jets in the other cases. 
 Table 3 shows that even when $\zeta$ is reasonably well constrained from PWN morphology, the $\alpha_{\rm pred}$ value(s) allowed by the magnetospheric emission models for  $\zeta=\zeta_{\rm PWN}$ can still span a wide range, which, however, shrinks substantially compared to what could be predicted from the magnetospheric emission models alone. 
The values of   $\alpha_{\rm pred}$ show that PWNe with  weaker toroidal components indeed tend to have smaller values of $\alpha$ (see, e.g., the B1509--58, Kes 75, and G11.2--0.3 PWNe in figure 2). 
The TPC model seems to be performing somewhat better compared to the OG model, in the sense that it has less trouble finding an allowed $\alpha_{\rm pred}$ for a given $\zeta_{\rm PWN}$. 
The next  step in this direction will be to increase the sample of PWNe with interpretable morphologies powered by $\gamma$-ray and radio pulsars. 
It would also be interesting to look at the properties of PWNe whose pulsars lack $\gamma$-ray  emission. 
With a sufficiently large sample, one could also start looking into correlations between $\alpha$ and other PWN parameters such as $\Gamma_{\rm i}$ and luminosity.

\begin{figure}
\centerline{\includegraphics[scale=0.34]{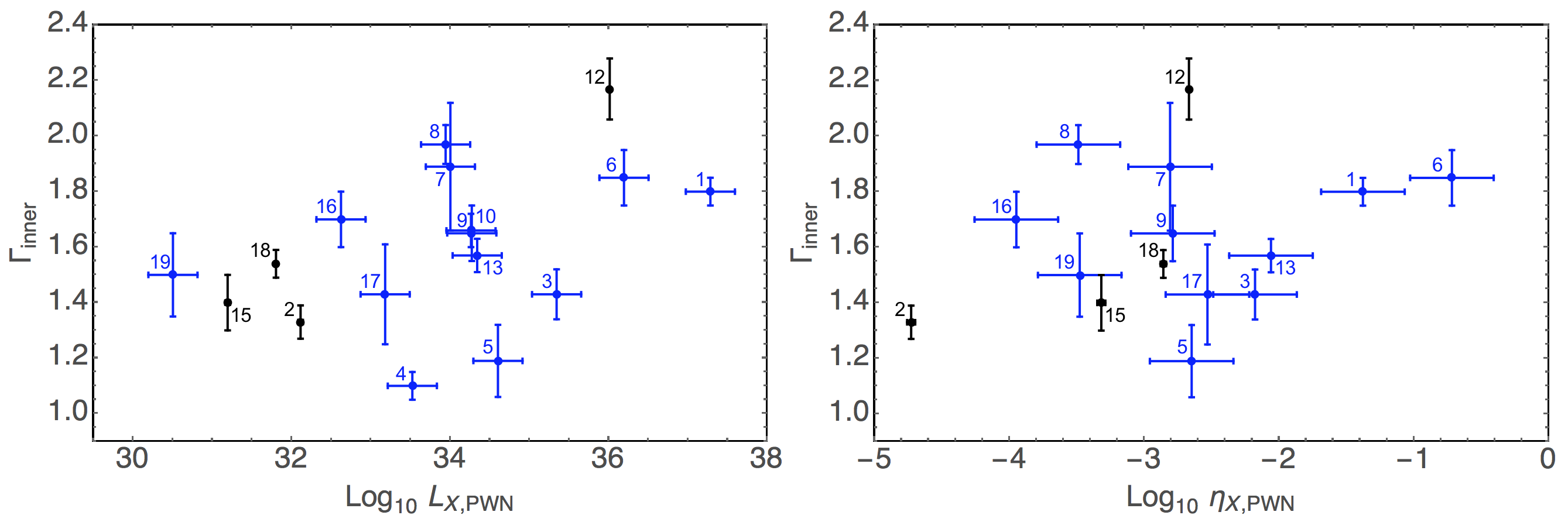}}
\vspace{-0.4cm}
\caption{ {\sl Left:} Photon index of the injection spectrum vs.\ X-ray luminosity 
  of the central regions of PWNe (from table 2 of \cite{Kargaltsev2008};
  the values for B0355, J1509, J1747, and J1741 are from \cite{Klingler2016a,Klingler2016b,Auchettl2015},
   respectively).   {\sl Right:}  Photon index of the injection spectrum vs.\ PWN radiative efficiency $\eta=L_X/\dot{E}$. 
   Pulsars/PWNe whose distances are known accurately (e.g., via parallax, and for N157B in the LMC) are shown in black. For the rest of the pulsars (shown in blue)  30\% uncertainties in the distances are assumed.}
 \label{fig:gammaLx}
\end{figure}

\section{Maximum Particle Energy in Pulsar Wind}
 The maximum energy of a synchrotron photon can be estimated by equating the synchrotron losses in a magnetic field $B$ to the energy gain in the electric field of strength $E\simeq B$ (for typical, highly-conducting astrophysical plasmas\footnote{Flares in the Crab PWN suggest that this condition may be violated in certain regions where reconnection occurs (see e.g., \cite{Kroon2016}).}
 $E<B$), which leads to the maximum synchrotron photon energy $\epsilon_{\rm \gamma,max}= \hbar  mc^3/e^2\sim100$ MeV.
 For electrons radiating close to this energy the acceleration must occur on a gyro-radius length scale,  which means that any stochastic acceleration (e.g., Fermi type; see \cite{Bykov2017}
  and references therein) would result in a substantially lower energy \cite{Lyutikov2010}.
 An estimate of the corresponding maximum electron energy requires knowing the magnetic field within the particle acceleration region, which is challenging because the acceleration region location and the acceleration process itself are still a matter of debate (see \cite{Lyutikov2016,Bykov2017}
 and references therein). 
   
On the other hand, if the energy for particle acceleration in PW ultimately comes from the potential drop available in a pulsar magnetosphere, $\Phi\sim(\dot{E}/c)^{1/2}$,  then the maximum  $\gamma_{\rm max}=\Phi e/m_e c^2\sim1\times 10^{10} (\dot{E}/10^{37}~{\rm erg/s})^{1/2}$.
 For the Crab, Vela, Geminga and the Guitar (B2224+65) pulsars, the corresponding values of $\gamma_{\rm max}$ are, respectively, $\sim7\times10^{10}$, $9\times10^{9}$, $6\times10^{8}$ and $1\times10^{8}$.  
The dependence on $\dot{E}$ suggests that by observing older (lower $\dot{E}$) pulsars in X-rays, one can probe the fraction of the available magnetospheric  potential drop   used to accelerate PW particles.
  The largest uncertainty in such estimates  comes from the PWN magnetic field  (or PW magnetization). Since it  is difficult to measure the field directly, it must be inferred from other PWN properties (e.g., from synchrotron brightness; see e.g., \cite{Pavlov2003}).
  This difficulty can possibly be circumvented for PWNe of supersonic pulsars which exhibit {\em misaligned outflows} (see below).

 The recently launched {\sl NuSTAR} observatory was able to detect and resolve 
 the  Crab (up to 78 keV; \cite{Madsen2015})
 and the B1509--58  (up to 40 keV; \cite{An2014})
 PWNe.  
 Such energies correspond to electron Lorentz factors of $\gamma_{\rm Crab}\sim  10^8$ and $\gamma_{\rm B1509}\sim 4\times10^8$ for plausible magnetic fields within these PWNe ($B_{\rm Crab}\sim300\ {\rm \mu G}$  \cite{Hester2008} and $B_{\rm B1509}\sim10{\rm \mu G}$ \cite{An2014}).  
    The lack of sensitive instruments with good angular resolution in the MeV range makes it challenging to determine, from synchrotron emission, if electrons with  even higher $\gamma$ are present.  Perhaps, the Crab PWN is the only PWN with  reliable MeV spectrum measurements showing that the (quiescent) synchrotron spectrum breaks sharply above $\sim$20 MeV \cite{Kuiper2001},
 which corresponds to $\gamma_{\rm Crab}\sim  4\times10^9$ for the assumed $B_{\rm Crab}\sim300\ {\rm \mu G}$. This falls substantially short of the $\gamma_{\rm max, Crab}\approx7\times10^{10}$ value based on the full magnetospheric potential drop, but it is close to the limit that is expected to be imposed by the synchrotron losses (see above). 
 
 The inverse Compton (IC) emission from  background photons up-scattered by PW electrons has been seen by HAWC (e.g., \cite{Abeysekara2017})  up to energies ($\epsilon_{\rm IC}$) of a few tens of TeV. 
   If the up-scattered photons come primarily from Cosmic Microwave Background (CMB),
  then   $\gamma\sim3\times 10^8 (\epsilon_{\rm IC}/60~{\rm TeV})^{1/2}(\epsilon_{\rm CMB}/6\times10^{-4}~{\rm eV})^{-1/2}$. This is close to the maximum $\gamma$ values inferred from synchrotron emission (see above) and may exceed  $\gamma_{\rm max}$ for older low-$\dot{E}$ pulsars which are believed to power relic PWNe (e.g., $\gamma_{\rm max}\sim6\times10^{8}$ for Geminga, whose relic PWN is likely associated with the  HAWC source seen above 20 TeV   \cite{Abeysekara2017}). 
 We note that although the TeV emitting particles can be produced earlier in  pulsar's life (when $\gamma_{\rm max}$ was higher), those particles that radiate at 60 TeV have cooling times of  around 10 kyr (e.g., Eqn.~4 in \cite{Kargaltsev2013}),
 which is significantly smaller than the 340 kyr spin-down age of Geminga. Therefore, these particles experienced  an accelerating potential corresponding to a nearly modern-day $\dot{E}$.

\begin{table*}
\scriptsize
\caption{
\footnotesize 
Comparison of various properties for four supersonic pulsars with misaligned outflows. For the apex stand-off distance $r_s$ we used the observed values for B0355+54 \cite{Klingler2016a} and B2224+65 \cite{vanKer2008Guitar}, and estimated $r_s$ from the pressure balance (for ISM number density $n=1$ cm$^{-3}$) 
  for J1509--5850 and J1101--6101. 
   The lower limits on the  Lorentz factors, $\gamma_{\rm esc}$, of electrons escaping into the ISM from the PWN apex  are estimated for  $B_{\rm ISM}=5~\mu$G  and synchrotron photon energy of 8 keV.   The  maximum Lorentz factors, $\gamma_{\rm max}$, are calculated from the full potential drop available in the pulsar magnetosphere (see text).   $B_{\rm apex}$ is the PWN magnetic field  near the PWN apex.    The  gyro-radii, $r_{\rm g, max}$, of  electrons  with Lorentz factor $\gamma_{\rm max}$ are calculated using  the upper limit on $B_{\rm apex}$ for the PWN magnetic field.  See  \cite{Chatterjee2004,Pavan2016,Klingler2016b,Klingler2016a} for the estimates of pulsar velocities, $v_{\rm PSR}$. }
 \centering
\setlength{\tabcolsep}{0.3pt}
\begin{tabular}{lcccccccc}
\hline\noalign{\smallskip}
Name \ \ \ \ \ \ \ \ & \ \ \ \ \ $\dot{E}$ \ \ \ \ \ & \ \ \ $P$ \ \ \ & \ \ $v_{\rm PSR}$ \ \ & \ \ $r_{\rm s}$ \ \ & \ \ $\gamma_{\rm esc}$ \ \  &  \ \ $\gamma_{\rm max}$ \ \  & \ \ $B_{\rm apex}$\ \   & \  $r_{\rm g, max}$ \  \\
  & \ ($10^{35}$ erg s$^{-1}$) \  & \ \ (ms)\ \  & \ \ (km s$^{-1}$)\ \  & \ ($10^{15}$ cm)\  & ($10^8$) & \ \ ($10^8$)\ \ & \ \ \ (${\rm \mu G}$)\ \ \ &  \ \ ($10^{15}$ cm) \ \  \\
\hline
B2224+65 (Guitar)             &  0.012  &  682  &  $\sim$1600  &  2.4  &  6  &  1  &  $<$125   &  1.4  \\
J1101--6101 (Lighthouse)  &  5.1     &  88.9  &  $\sim$2000  &  6  &  6  &  39  &  $<$57    &  117  \\
J1509--5850                      &  14      &  62.8  &  $\sim$800    &  10   &  6  &  24  & $<$34    &  120  \\
B0355+54 (Mushroom)$^{\dagger}$                       &  0.45  &  156   &  $61^{+12}_{-9}$  &  54   &  6  &  7.1  &  $<$6   &  202 \\
\noalign{\smallskip}\hline
\end{tabular}\\
\vspace{-0.3cm}
\label{tbl-pulsars2}
 \end{table*}

  An alternative approach to the estimation of $\gamma_{\rm max}$ has become possible with the discovery of misaligned outflows produced by a few supersonically moving pulsars (B2224+65, J1101--6101, J1509--5850, and possibly B0355+54; see \cite{Kargaltsev2017}
  and references therein).   
 These puzzling structures can be explained by the kinetic leakage of the most energetic particles near the apex of the bow shock, which becomes possible when the bow shock stand-off distance, $r_{\rm s} \simeq (\dot{E}/4\pi c  m_H n  v_p^2)^{1/2}$ ($n$ is the ISM baryon number density; see e.g., \cite{Kargaltsev2017}),
 is comparable to the  gyro-radius, $r_g=\gamma m_e c^2/eB_{\rm apex}$, in the PWN magnetic field $B_{\rm apex}$ near the bow-shaped PWN apex \cite{Bandiera2008}.
 This condition can be used to put a lower limit on the energies of particles  escaping into the ISM, if the ISM magnetic field is known (or assumed).  Since   supersonically moving pulsars with ages $\gtrsim 10$ kyr  are expected to leave their host SNRs and move in a relatively unperturbed
   ISM with typical $B_{\rm ISM}\simeq 5\ {\rm \mu G}$ \cite{Opher2009},
    the Lorentz factor of the electrons that escaped into the ISM can be estimated as   $\gamma_{\rm esc} \gtrsim 6\times10^{8}(\epsilon/8\,{\rm keV})^{1/2}(B_{\rm ISM}/5\,\mu{\rm G})^{-1/2}$, where $\epsilon$ is the observed energy of synchrotron photon (8 keV is the upper boundary of the {\em CXO} band).   From the escape condition, $r_g \gtrsim r_{\rm s} $,  one can also set an upper limit on the PWN field near the apex,
 $B_{\rm apex}\lesssim34(\epsilon/1\,{\rm keV})^{1/2}(B_{\rm ISM}/5\,\mu{\rm G})^{-1/2}(r_{\rm s}/10^{16}~{\rm cm})^{-1}$~$\mu$G (here we adopted 1 keV as the lower  boundary of the {\sl CXO} energy band because the  typical PWN spectra are  substantially absorbed below this energy).  As one can see from table 4, for B0355+54 the upper limit on  $B_{\rm apex}$ approaches the ISM field and therefore provides a good constraint suggesting that  $B_{\rm apex}$ is a few $\mu$G.   
   Table 4 also shows a comparison of  $\gamma_{\rm max}$ and  $\gamma_{\rm esc}$ for four supersonic pulsars with misaligned outflows.  In the case of the Guitar nebula (PSR B2224+65), the energies of X-ray radiating electrons  in the misaligned outflow appear to exceed the energy available in the magnetospheric potential drop (due to the space limitations, we refer the reader to  \cite{Bykov2017,Petri2016}
    for  discussion and possible explanation).  If a substantial number of particles are able to reach $\gamma_{\rm max}$ (a plausible situation for the old PSR B0355+54 given that it apparently happens in the somewhat older PSR B2224+65),  they should be able to leak into the ambient  ISM very easily since the gyro-radii  of electrons with Lorentz factors $\gamma_{\rm max}$ 
    would  greatly exceed the  stand-off distance $r_s$.

\section{Summary}

We discussed the observational properties of PWNe, linking them to the injected (post-TS) electron SED and  parameters of pulsar magnetospheres.  The main findings can be summarize as follows.

\begin{itemize}
\item The photon indices of PWNe measured in X-rays for the innermost resolved regions of bright, extended PWNe vary in the range of  $\Gamma_i=1.1-2.2$ with the majority (80\%) being in the  $\Gamma_i=1.3-2.0$ range. The corresponding ranges of the uncooled electron SED slopes, $p_i=2\Gamma_i-1$, are  1.2--2.9 and 1.8--2.5, respectively. Effects of cooling are obvious in many high-resolution spectral maps of PWNe, hence inferring $p_i$ from  the spatially-averaged  $\Gamma$  measured from a large PWN region  does not provide accurate diagnostics of  particle SED injected at the TS.  The spectra of the inner regions of the tori are similar to those of the inner jets,
  in the same PWN.
\item Measuring the viewing angle  $\zeta$ from approximately axisymmetric  morphologies of PWNe with tori and/or jets allows one to test pulsar magnetosphere models which are now capable of predicting $\zeta$ and the magnetic inclination $\alpha$ based on $\gamma$-ray and radio light curves. 
   A correlation between  $\alpha$ and PWN properties  can be expected from general principles and is supported by recent numerical modeling  \cite{Buhler2016}.
 We used the $\alpha$ values predicted by magnetospheric models  to 
  look for a correlation between the PWN spatial morphologies, luminosities, or spectra.  We find some observational support for a hypothesis that pulsars with smaller $\alpha$ power PWNe with relatively weaker toroidal components (compared to their jet components).
 \item  The capabilities of modern X-ray and TeV $\gamma$-ray observatories allow one to infer (under some assumptions) PW particle energies. These energies approach, and in one case (B2224+65) exceed, the full potential drop available in pulsar magnetosphere. 

\end{itemize}

\noindent{\sl Acknowledgements:}  We thank  Andery Bykov and Demos Kazanas for stimulating discussions. We acknowledge partial support from  {\sl CXO} awards  GO3-14082,  GO3-14084, GO3-14057 and AR5-6006.

\section*{References}

\vfill
\begingroup
\let\clearpage\relax
\include{adsjournalnames}
\bibliographystyle{iopart-num}
\bibliography{bibliography.bib}
\endgroup

\end{document}